\documentclass[journal]{IEEEtran}

\usepackage{graphicx}
\usepackage{amsmath}
\usepackage{amssymb}
\usepackage{mathrsfs}
\usepackage{cite}

\newcommand{\rmi}{ \mathrm{i} }

\renewcommand\deg{^\circ}

\usepackage{color}
\graphicspath{{./Figures/}}

\usepackage{nicefrac}
\usepackage{xspace}
\usepackage{color}
\usepackage[normalem]{ulem}

\newcommand{\textoutr}[1]{}

\newcommand{\fst}{FeSe$_x$Te$_{1-x}$ }
\newcommand{\rhoff}{\rho_{\it ff}}

\begin{document}

\title{Pinning, flux flow resistivity and anisotropy of Fe(Se,Te) thin films from microwave measurements through a bitonal dielectric resonator}

\author{N.~Pompeo,~\IEEEmembership{Senior Member,~IEEE,}
        A. Alimenti,~\IEEEmembership{Student Member,~IEEE,} 
        K. Torokhtii,~\IEEEmembership{Member,~IEEE,} 
        G. Sylva, 
        V. Braccini,
        E.~Silva,~\IEEEmembership{Senior Member,~IEEE}% 
\thanks{A. Alimenti, N. Pompeo, E. Silva and K. Torokhtii are with the Department of Engineering, Universit\`{a} Roma Tre, 00146 Roma, Italy. Corresponding author: N. Pompeo; e-mail: nicola.pompeo@uniroma3.it.}% 
\thanks{V. Braccini and G. Sylva are with CNR-SPIN Genova, C.so F. M. Perrone, 24, 16152 Genova, Italy.}% 
}

{}

\maketitle

\begin{abstract}
We report on the anisotropy of the vortex motion surface impedance of a \fst thin film grown on a CaF$_2$ substrate.
The dependence on the magnetic field intensity up to 1.2 T and direction, both parallel and perpendicular to the sample $c$-axis, was explored at fixed temperature at two distinct frequencies, $\sim16\;$GHz and $\sim27\;$GHz, by means of bitonal dielectric resonator.
The free flux flow resistivity $\rhoff$ was obtained by exploiting standard models for the high frequency dynamics, whereas the angle dependence was studied in the framework of the well known and widely used Blatter-Geshkenbein-Larkin (BGL) scaling theory for anistropic superconductors.
Excellent agreement with the scaling law prescription by the fluxon flux flow resistivity was obtained.
From the scaling analysis, a low-field mass anisotropy $\sim1.8$ was obtained, well within the value ranges reported in literature. The angular dependence of the pinning constant suggests that pinning is dominated by random, isotropic point pins, consistently with critical current density measurements.
\end{abstract}


\IEEEpeerreviewmaketitle

\section{Introduction}
\label{sec:intro}

\IEEEPARstart{S}{ince} their discovery, Iron-Based Superconductors (IBS) have attracted an intense research effort \cite{Hosono2017} due to their intriguing physics and perspective applications, thanks to their relatively high critical temperature $T_c$ and irreversibility field $H_{irr}$, low anisotropy, reduced impact of grains misalignment on the critical current $J_c$.
The improvement of the pinning properties, with perspective transfer to the coated conductor technology \cite{Xu2017}, requires a deep comprehension and subsequent control of the material properties, with special care to the pinning landscape and to the overall anisotropy. 
The reported anisotropy \cite{Hosono2017} of \fst is in general an advantage for the applications, although the multigap nature of IBS makes the evaluation of the anisotropy intricate.

The conventional definition of the (uniaxial) mass anisotropy is $\gamma=\sqrt{m_c/m_{ab}}$, where $m_c$ and $m_{ab}$ are the electronic masses along the $c$-axis and the $ab$-planes respectively. 
The often used critical current density anisotropy $\gamma_J=J_{c}^{\parallel ab}/J_{c}^{\parallel c}$, relevant for the assessment of a material performance in d.c. applications, depends not only on the mass anisotropy but also on the dimensionality of the pinning centers: as soon as the latter include 3D or elongated (1D, 2D) pins, $\gamma_J\neq\gamma$.
In single-band superconductors the same anisotropy value as the mass anisotropy is obtained when evaluated as coherence length, penetration depth and lower/upper critical field ratios ($\gamma_\xi=\xi_{ab}/\xi_c$, $\gamma_{\lambda}=\lambda_c/\lambda_{ab}$, $\gamma_{H_{c1}}=H_{c1}^{\parallel c}/H_{c1}^{\parallel ab}$, $\gamma_{H_{c2}}=H_{c2}^{\parallel ab}/H_{c2}^{\parallel c}$, respectively) \cite{Tinkham1996book}. 
On the other hand, in multigap superconductors the interplay of different energy and length scales with potentially different temperature dependences, with the addition of Pauli limited upper critical fields \cite{Lei2010}, makes the above different declinations of the anisotropy not necessarily equivalent. Nevertheless it has been shown \cite{Kidszun2011, Hanisch2011} that the Blatter-Geshkenbein-Larking (BGL) theory for the scaling of anistropic quantities \cite{Blatter1992}, conceived for single-gap superconductors, can be applied in multigap IBS. The so-obtained \emph{effective} anisotropies may develop a temperature dependence, yielding generally different values among the various declinations \cite{Iida2013, Zhaofeng2016, Grimaldi2018, Leo2019}. 

It is therefore of interest to address the issue from a yet different point of view.
It has been demonstrated that through high frequency measurements of the vortex motion resistivity $\rho_{vm}$ it is possible to extract the mass anisotropy from the material intrinsic quantity, i.e. pinning-independent, flux flow resistivity $\rhoff$ \cite{Pompeo2013,Bartolome2019,Pompeo_2020,PompeoLTP2020}. Moreover, within the same measurements, taking advantage of the complex nature of a.c. quantities, also the pinning anisotropy can be evaluated. 
Hence, we report here a preliminary study on the anisotropy of a \fst thin film grown on CaF$_2$ based on multifrequency microwave measurements of its vortex motion resistivity $\rho_{vm}$. Measurements are performed by means of a bitonal dielectric resonator (DR) simultaneously operating at $\sim16\;$GHz and $\sim27\;$GHz, used with the surface perturbation method, with static magnetic fields up to 1.2 T applied both parallel and perpendicular to the sample normal.

This paper is structured as follows: in Section \ref{sec:experiment} the experimental technique is described. In Section \ref{sec:model} we briefly recall the high frequency electrodynamic model, which includes the vortex motion parameters of interest, and the BGL scaling theory applied to the study of the angular dependence. In Section \ref{sec:results}, the measurements will be presented and analysed.
\section{Experimental setup}
\label{sec:experiment}

The superconductor surface impedance $Z$
is measured by means of a dielectric-loaded cylindrical electromagnetic (e.m.) resonator within the surface perturbation method \cite{Pompeo2014}.
The complex, frequency ($\nu$) dependent, scattering coefficient $S_{21}(\nu)$ of the resonator is measured through a Vector Network Analyser (VNA) an then fitted, taking into account various non-idealities including line calibration issues \cite{Alimenti2019a}, to extract the quality factor $Q$ and resonant frequency $\nu_0$ at resonance.
Through standard e.m. theory \cite{Collin1992}, the magnetic field $H$ response of the sample in terms of its surface impedance $Z=R+\rmi X$ is computed from $Q$ and $\nu_0$:
\begin{eqnarray}
\label{eq:QZ}
\Delta R(H)+\rmi\Delta X(H)=G\Delta\frac{1}{Q(H)}-\rmi2G\frac{\Delta\nu_0(H)}{\nu_0(0)}
\end{eqnarray}
where $G$ is a mode dependent geometrical factor determined through numerical simulations and ${\Delta A(H)=A(H)-A(0)}$ represents the variation of the quantity $A$ induced by the application of a static magnetic field $H$ with respect to the zero field value $A(0)$, keeping fixed the temperature $T$ and the magnetic field direction $\theta$ relative to the sample normal (i.e. superconductor $c$-axis).
The resonator does not show any magnetic signal.
In electromagnetically thin samples, i.e. having thickness $t_s\ll \min(\lambda, \delta)$ where $\lambda$ and $\delta$ are the London penetration and skin depths, respectively, one exploits the so-called thin film approximation $Z\approx\tilde\rho/t_s$ (where $\tilde\rho$ is the sample complex resistivity, working in the local limit) \cite{PompeoImeko2017a}. Moreover, far from the transition line $H_{c2}(T)$, field pair-breaking effects can be neglected and $\Delta\tilde\rho(H)=\rho_{vm}(H)$, so that ultimately the desired $\rho_{vm}(B)=\Delta Z(H) t_s$, where the London limit $B=\mu_0 H$ is considered. 

The bitonal resonator here used has been specially designed in order to operate at two distinct resonant frequencies ${\nu_{01}\simeq16.4\;}$GHz and $\nu_{02}\simeq26.6\;$GHz, corresponding to the transverse electric modes TE$_{011}$ and TE$_{021}$, selected for 
their high $Q$ delivering high sensitivity, and the planar, circular current patterns that they induce on the sample surface \cite{Pompeo2014, Alimenti2019a}. Then, for a given $T$, $H$ and $\theta$, $\rho_{vm}(\nu)$ with ${\nu=\{\nu_{01}, \nu_{02}\}}$ can be measured. 
The measurement cell is placed inside a He-flow cryostat and within the bore of a conventional rotating electromagnet, capable of supplying magnetic fields $\mu_0H\leq1.2\;$T with varying direction with respect to the resonator axis. 

The \fst sample under investigation is $t_s=240\;$nm thick and was deposited on a 7~mm side square CaF$_2$ single crystal substrate by Pulsed Laser Deposition starting from a FeSe$_{0.5}$Te$_{0.5}$ target \cite{Palenzona2012}. The rotational symmetry of the resonator is preserved by covering the square sample with a thin metal mask having a circular hole in it.
The sample has ${T_c\simeq18\;}$K and a normal state resistivity ${\rho_n=(3.0\pm0.2)\cdot10^{-6}\;\rm{\Omega m}}$ \cite{Pompeo2020a}, in good agreement with the values in similar samples \cite{Palenzona2012,PompeoEUCAS2019} or other single/poly- crystals of similar composition \cite{Okada2015,Li2015}.

\begin{figure}[htbp]
\centering
\includegraphics[width=0.8\columnwidth]{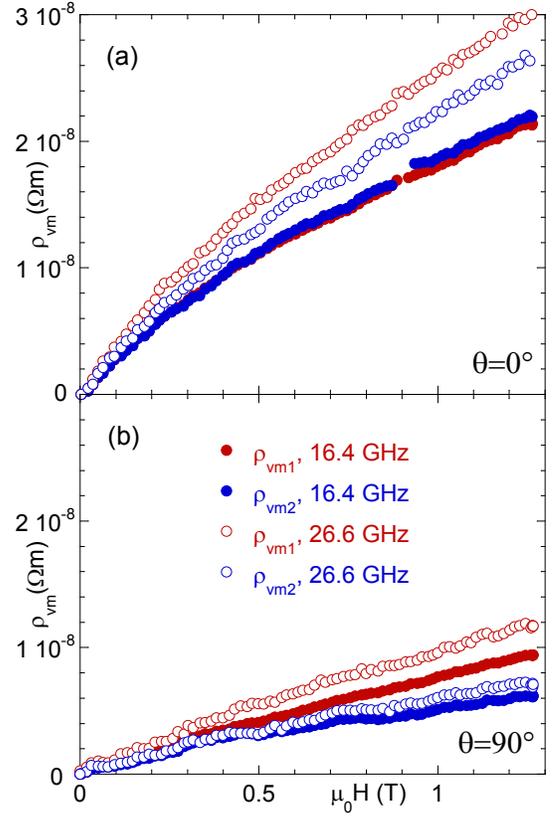}
\caption{Vortex  motion resistivity at 13 K for $H$ parallel ($\theta=0\deg$) and perpendicular ($\theta=90\deg$) to the sample $c$-axis, panel (a) and (b) respectively.}
\label{fig:rhovm}
\end{figure}

In Fig. \ref{fig:rhovm} we report the experimental data ${\rho_{vm}=\rho_{vm1}+\rmi\rho_{vm2}=}\Delta Z t_s$ at the two frequencies as a function of the magnetic field applied parallel (Fig.\ref{fig:rhovm}a) and perpendicular (Fig.\ref{fig:rhovm}b) to the c-axis. In order to discuss the vortex motion parameters we shortly present the theoretical framework in the next Section.
\section{The models}
\label{sec:model}

The vortex motion resistivity $\rho_{vm}$ is studied within the standard models for vortex motion  \cite{Gittleman1966, Coffey1991a, Brandt1992} (see \cite{Golosovsky1996, Pompeo2008} for short reviews). On quite general grounds, the following expression can be written down:
\begin{eqnarray}
\label{eq:rhovm}
\rho_{vm}&=\rhoff\frac{\chi+\rmi\nu/\nu_c}{1+\rmi\nu/\nu_c}
\end{eqnarray}
The scale factor $\rhoff=\alpha\rho_n B/B_{c2} $ is the free flux flow resistivity, 
where $\alpha=1$ in the basic Bardeen Stephen (BS) \cite{Bardeen1965} model, 
and is connected with the quasi-particles scattering time and density of states in the vortex cores \cite{Caroli1964} thus yielding information about the mass anisotropy $\gamma$.
The creep factor $\chi\in[0,1]$ weights the thermal depinning phenomena.
The characteristic frequency $\nu_c$ is of special importance in assessing the high frequency dissipation levels of superconductors in the mixed state, since it marks the transition from a low-loss regime at $\nu\ll\nu_c$ to a high-loss regime at $\nu\gg\nu_c$ \cite{Alimenti2020, Alimenti2020a}. 
When $\chi=0$ no creep effects are present, Eq. \eqref{eq:rhovm} yields the Gittleman and Rosenblum (GR) expression \cite{Gittleman1966}, and $\nu_c\rightarrow\nu_p=k_p\rhoff/(2\pi B \Phi_0)$ the well-known (de)pinning frequency. In the latter expression, $\Phi_0$ is the flux quantum and $k_p$ the pinning constant, also known as Labusch parameter, a measure of the pinning well steepness/curvature.
If on one hand $\rhoff$ is a window into the intrinsic material properties, the pinning-related quantities $\nu_c$, $k_p$ and $\chi$ are ``extrinsic'' since they depends on the pinning phenomena and thus on the defect distribution in the material.

With measurements at two frequencies as presented here, all the parameters in Eq. \eqref{eq:rhovm} can be directly obtained, overcoming the large uncertainties in single frequency measurements \cite{Pompeo2008}.

The anisotropy is derived as follows. The Blatter-Geshkenbein-Larkin (BGL) scaling \cite{Blatter1992} theory, for high-$\kappa$ uniaxially anisotropic superconductors in the London limit and with 0D pinning points only, 
states that for a field and angle dependent observable $\mathcal{Q}(H, \theta)$ 
the field and angular dependences combine into an effective field $H\epsilon(\theta)$ such that:
\begin{subequations}
\begin{eqnarray}
\mathcal{Q}(H, \theta)&=&s_{\mathcal{Q}}(\theta)\mathcal{Q}(H\epsilon(\theta))\\
\epsilon(\theta)&=&(\gamma^{-2}\sin^2\theta+\cos^2\theta)^{1/2}
\end{eqnarray}
\label{eq:scaling}
\end{subequations}
where $\epsilon(\theta)$ is the so-called anisotropy factor and $s_{\mathcal{Q}}(\theta)$ is a $\mathcal{Q}$-dependent, a-priori known, multiplicative factor.
In our high frequency study, by measuring the intrinsic (i.e. independent from pinning) quantity $\rhoff$ for different angles $\theta$, the angular scaling allows to extract the material anisotropy $\gamma$.
The scaling rule can be also tested in purely pinning-related quantities like $k_p$ and $\chi$: if it applies, it means that only 0D pins are involved, otherwise other - extrinsic - sources of anisotropy (i.e. extended defects of 1D, 2D and/or 3D dimensionality) are present, similarly to the effects reported in nanostructured YBa$_2$Cu$_3$O$_{7-\delta}$ \cite{Pompeo2013, Bartolome2019}.

\section{Results and discussion}
\label{sec:results}

We focus here on a complete set of measurements taken at $T=13\;$K, deferring a complete temperature study to a future study. In Figure \ref{fig:rhovm} we presented the data for 
$\rho_{vm}(H)$, in terms of its real $\rho_{vm1}$ and imaginary $\rho_{vm2}$ parts, at $\nu_{01}=16.4\;$GHz and $\nu_{02}=26.6\;$GHz, for $\theta=0\deg$ ($H$//$c$-axis) and $\theta=90\deg$ ($H$//$ab$-planes).
Some standard features can be recognized: a slight downward curvature\footnote{A linear $H$-dependence would call for a Bardeen-Stephen flux-flow resistivity and field-independent flux-creep and flux-pinning, see Eq. \eqref{eq:rhovm}.};
similar real and imaginary parts, hinting that $\nu_c$ is not far from the measurement frequencies; smaller resistivity in the $H$//$ab$ orientation with respect to $H$//$c$, pointing to an anisotropic behaviour whose origin (intrinsic, pinning-related, or both) remains at this stage to be assessed. 

\begin{figure}[htbp]
\centering
\includegraphics[width=0.8\columnwidth]{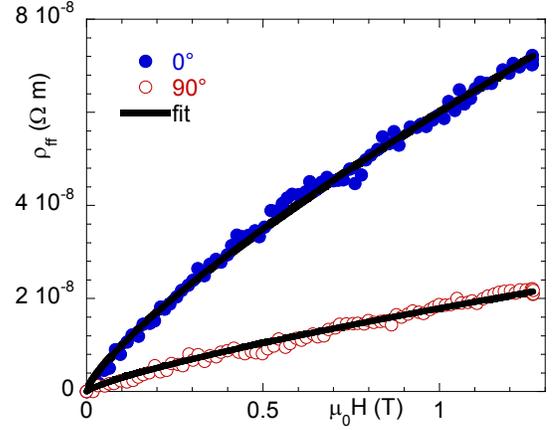}
\caption{Flux flow resistivity at $13\;$K. Thick lines: fits with $H^\beta$, $\beta=0.78$.}
\label{fig:rhoff}
\vspace{-2mm}
\end{figure}

\begin{figure}[htbp]
\centering
\includegraphics[width=0.8\columnwidth]{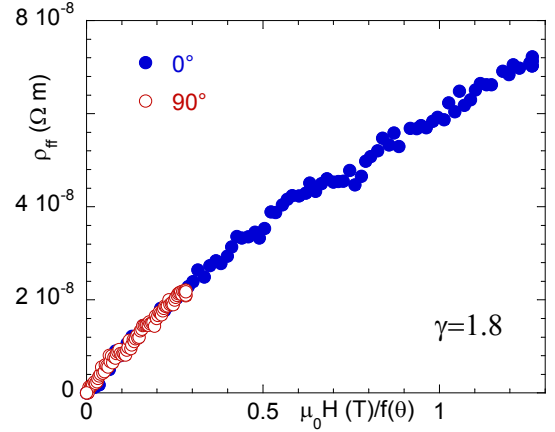}
\caption{Scaled flux flow resistivity at $13\;$K.}
\label{fig:rhoff_scaled}
\end{figure}

The vortex motion parameters ($\rhoff$, $\nu_c$ and $\chi$) - for the two field directions - are extracted by analytically inverting Eq. \eqref{eq:rhovm} against the experimental curves.
The flux flow resistivity $\rhoff(H)$ is shown in Fig. \ref{fig:rhoff}: it can be seen that it exhibits a slight down-curvature, which can be fitted with the power law $H^{\beta}$ with the same $\beta=0.78$ for both directions, and a much distinct amplitude for the two field directions. 
We now proceed with the anisotropy study.
By a proper scale factor $f(\theta=90^\circ)=4.5$ ($f(\theta=0^\circ)=1$ by definition) it is possible to superimpose the curve $\rhoff(H, \theta=90^\circ)$ vs $H/f(\theta=90^\circ)$ over the curve $\rhoff(H, \theta=0^\circ)$ vs $H$ (see Fig. \ref{fig:rhoff_scaled}).
It can be seen that a very good scaling can be obtained, 
experimentally demonstrating that the scaling takes place in our \fst sample.
Before interpreting the data, we recall that our experimental setup 
induces on the sample circular currents with angle-dependent Lorentz forces acting on the fluxons. This technique-specific factor leads to an effective scaling function \cite{Pompeo2013,Pompeo2015,Pompeo2018a}:
\begin{subequations}
\begin{eqnarray}
f(\theta)&=\epsilon(\theta)^{-1}f_L(\theta)^{1/\beta}\\
f_L(\theta)&=\frac{\gamma^{-2}\sin(\theta)^2+\cos(\theta)^2}{\frac{\gamma^{-2}}{2}\sin(\theta)^2+\cos(\theta)^2}
\end{eqnarray}
\label{eq:fscaling}
\end{subequations}
where $f_L(\theta)$ is due to nonhomogenous Lorentz force and $\beta$ is the exponent for the $\rhoff$ field dependence previously determined. From the above Eq.s \eqref{eq:scaling} and \eqref{eq:fscaling},  for $\rhoff$ with $s_{\rhoff}=1$ \cite{Blatter1994}, the scaling yields an intrinsic anisotropy $\gamma=1.8$, the first important result of this work and, to the best of our knowledge, the first determination of the intrinsic anisotropy in \fst by microwave studies. 
It is worth stressing that it has been shown that the proposed analysis, even if based on the scaling of two curves only instead of a wider scaling of several curves for several $\theta$ values, is reliable \cite{Bartolome2019, Pompeo_2020}.  
The obtained $\gamma$ belongs to the range $\gamma_{H_{c2}}=1 - 2.5$ found in samples of similar composition, both single crystals and films \cite{Tarantini2011,Bellingeri2014, Yuan2016, Zhaofeng2016, Leo2019, Grimaldi2019}.

\begin{figure}[htbp]
\centering
\includegraphics[width=0.8\columnwidth]{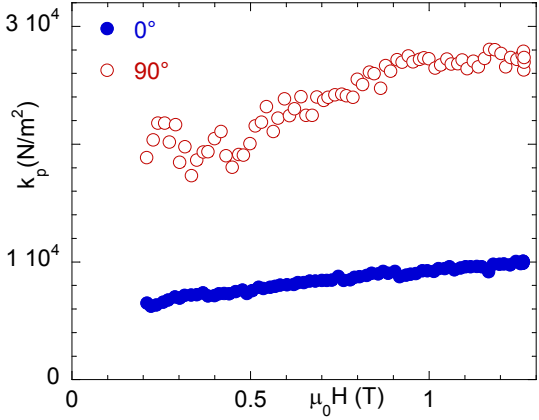}
\caption{Pinning constant at $13\;$K. Low field values are affected by inversion inaccuracies and thus discarded.}
\label{fig:kp}
\end{figure}

\begin{figure}[htbp]
\centering
\includegraphics[width=0.8\columnwidth]{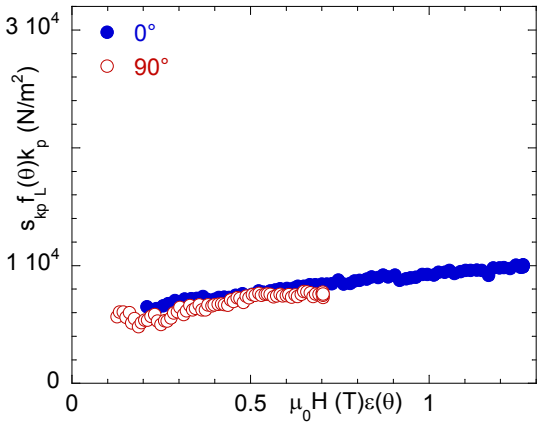}
\caption{Scaled pinning constant at $13\;$K.}
\label{fig:kpscaled}
\end{figure}

We now consider the anisotropy of the pinning constant $k_p$, as obtained from Eq.\eqref{eq:rhovm} using the relations between $\nu_c$ and $\chi$ within the theory detailed in \cite{Coffey1991a}. 
It is worth recalling that the dual frequency measurements here exploited allow the correct determination of the $k_p(H)$ curve, whereas single-frequency measurements would yield large uncertainties in $k_p$ and therefore a reduced reliability of the subsequent scaling \cite{Pompeo2008}.
The $k_p(H)$ curves for the two field directions are plotted in Fig. \ref{fig:kp}.
It can be seen that $k_p(90\deg)>k_p(0\deg)$: whether this is due to more effective pinning along the $ab$ planes, for example due to the so-called intrinsic pinning caused by the layered structure, or only a straightforward effect of the mass anisotropy, can be quantitatively checked as follows. 
Since $\gamma$ has been already obtained from $\rhoff$, the $k_p(H,\theta)$ curves can be plotted as scaled quantities.
First, the Lorentz-force spurious angular dependence must be removed by computing $k_p(H,\theta)f_L(\theta)$ \cite{Pompeo2013, Pompeo2015}. Then, the resulting quantity can be scaled according to Eq. \eqref{eq:scaling} with $s_{k_p}(\theta)=\epsilon(\theta)$ \cite{Pompeo2015}. The result is reported in Fig. \ref{fig:kpscaled}:
it can be seen that the scaled $k_p(90\deg)$ superimposes well over $k_p(0\deg)$. 
The excellent scaling indicates that pinning is dominated by point pins.
Thus the present \fst sample grown on CaF$_2$ does not present any visible sign of extended pins (neither elongated defects nor the so-called ``intrinsic'' pinning to the planes of the layered superconductor). Whether this result remains true also at different temperatures will be assessed in an extended, future work. 
This further confirms and validates the results reported in \cite{Braccini2013}, where angular dependent $J_c(\theta)$ on similar films provided flatter trends with respect to those grown on SrTiO$_3$ substrates. 

Summarizing, we have presented dual frequency microwave measurements of the vortex motion resistivity $\rho_{vm}$ on a \fst thin film grown on a CaF$_2$ substrate  with a static magnetic field up to 1.2 T applied parallel and perpendicular to the superconductor $c$-axis in order to assess its mass anisotropy $\gamma$ and pinning anisotropy. By standard vortex motion models and resorting to the BGL scaling, which was successfully applied to our data, we have obtained $\gamma=1.8$ at $T=13\;$K, in agreement with values reported in literature. Moreover, the pinning constant angular scaling showed that the pinning is remarkably isotropic, being dictated by the mass anisotropy only. This is an indication that the pinning landscape consists of random point pins only, apparently devoid of elongated pinning defects and of effective planar ``intrinsic'' pinning. Further works in extended temperature and angle ranges are planned.

\section*{Acknowledgments}
Work partially supported by MIUR-PRIN project “HiBiSCUS" - grant no. 201785KWLE.
The authors acknowledge A. Provino and P. Manfrinetti for the target preparation, and M. Putti for scientific discussion.

\bibliographystyle{IEEEtran}
\bibliography{MyCollection,MyBSTcontrol}

\end{document}